\documentclass[aps,prb,reprint]{revtex4-1} 
\usepackage[a4paper, total={7in, 10in}]{geometry}
\usepackage{graphicx} 
\usepackage{natbib} 
\usepackage{amsmath} 
\usepackage{amssymb}
\usepackage{amsmath}
\usepackage{graphicx}
\usepackage[english]{babel}
\usepackage[T1]{fontenc}
\usepackage[latin1]{inputenc}
\usepackage[pdftex,colorlinks=true,pdfstartview=FitH,linkcolor=blue,citecolor=blue,urlcolor=blue]{hyperref}
\usepackage{ctable}
\usepackage{subcaption}
\usepackage{color}
\usepackage{placeins}

\definecolor{azul}{rgb}{0.0, 0.0, 1.0}

\newcommand{\figreffc}[1]{Figure~\ref{#1}{\color{azul} \,c}}
\newcommand{\figreff}[1]{Figure~\ref{#1}}
\newcommand{\ecreff}[1]{Equation~{\color{blue}(\ref{#1})}}

\newcommand{\figref}[1]{Fig.~\ref{#1}}
\newcommand{\tableref}[1]{Tab.~\ref{#1}}
\newcommand{\figrefa}[1]{Fig.~\ref{#1}{\color{azul} \,a}}
\newcommand{\figrefb}[1]{Fig.~\ref{#1}{\color{azul} \,b}}
\newcommand{\figrefc}[1]{Fig.~\ref{#1}{\color{azul} \,c}}
\newcommand{\figrefd}[1]{Fig.~\ref{#1}{\color{azul} \,d}}
\newcommand{\ecref}[1]{Eq.~{\color{blue}(\ref{#1})}}
\newcommand{\xtx}[1]{\mathrm{#1}}
\newcommand{\mum}{\,$\mu$m}
\newcommand{\mus}{\,$\mu$s}
\newcommand{\nm}{\,nm}
\newcommand{\ns}{\,ns}

\newcommand{\kHz}{\,kHz}
\newcommand{\K}{\,K}
\newcommand{\Hz}{\,Hz}
\newcommand{\MHz}{\,MHz}
\newcommand{\simi}{$\sim$\,}
\newcommand{\mW}{\,mW}
\newcommand{\eV}{\,eV}
\newcommand{\meV}{\,meV}
\newcommand{\mueV}{\,$\mu$eV}

\begin{document}

\title{\textbf{Three dimensional trapping of light with light in semiconductor planar microcavities}}


\author{S. Anguiano$^{1}$, A. A. Reynoso$^{1}$,  A. E. Bruchhausen$^{1}$,  A. Lema\^itre$^2$, J. Bloch$^3$, and A. Fainstein$^{1 }$}
\affiliation{$^1$Centro At\'omico Bariloche \& Instituto Balseiro, CNEA, CONICET, 8400 San Carlos de Bariloche, R\'io Negro, Argentina}
\affiliation{$^2$Centre de Nanosciences et de Nanotechnologies, CNRS., Univ. Paris-Sud, Universite\'e Paris-Saclay, 91120 Palaiseau, France}
\affiliation{$^3$Centre de Nanosciences et de Nanotechnologies (C2N), Avenue de la Vauve, 911200 Palaiseau, France}

\begin{abstract}
When light is confined in all three directions and in dimensions of the order of the light wavelength, discretization of the photon spectra and distinctive phenomena occur, the Purcell effect and the inhibition of emission of atoms being two paradigmatic examples. Diverse solid-state devices that confine light in all three dimensions have been developed and applied. Typically the confinement volume, operating wavelength, and quality factor of these resonators are set by construction, and small variations of these characteristics with external perturbations are targeted for applications including light modulation and control. Here we describe full 3D light trapping, that is set and tuned by laser excitation in an all-optical scheme. The proposed device is based on a planar distributed Bragg reflector GaAs semiconductor microcavity operated at room temperature. Lateral confinement is generated by an in-plane gradient in the refractive index of the structure's materials due to localized heating, which is in turn induced by carriers photoexcited by a focused laser. Strong three dimensional trapping of light is evidenced by the laser-induced changes on the spectral, spatial, and k-space distribution of the emission. The dynamics of the laser induced photonic potential is studied using modulated optical excitation, highlighting the central role of thermal effects at the origin of the observed phenomena.
\end{abstract}

\maketitle
\begin{center}

\end{center}

Resonant cavities are pervasive to many different natural phenomena and man made devices. Standing waves confined between reflective walls build up, localizing the wave energy and enhancing their interaction with other physical degrees of freedom. They are ubiquitous in the domain of optics, for example as the basic feedback mechanism of lasers \cite{VCSEL,Tamm}, or in the most sensitive displacement sensors, such as the Fabry Perot cavities used for the detection of gravitational waves at LIGO \cite{LIGO}. When the confinement is strong and in all three dimensions, a discretization of part of the photon spectra  can occur \cite{Pilares,SizeDependence,Tamm,Paulo}, leading to distinctive phenomena as for example the inhibition or enhancement of the emission of atoms \cite{PurcellOriginal,Kleppner,Yablonovitch}, and the Bose-Einstein condensation of strongly coupled matter and light particles (cavity polaritons) \cite{Balili,Paulo}. In the NIR-visible spectral range, such three-dimensional photon trapping typically requires the development of confining structures based on nano- and micro-fabrication methods. Diverse techniques have been developed and applied, including the use of local defects in periodic photonic bandgap free-standing membranes \cite{Akahane}, the introduction of traps in planar distributed Bragg reflector (DBR) cavities through defects such as GaAs droplets \cite{Droplets} or lateral pattering \cite{Tamm,Mesas2006,Paulo}, or their full 3D structuring into microdisks and pillar resonators based on finely tuned ion etching methods \cite{Pilares,SizeDependence,McCall}. These techniques are already mature and robust, leading to high quality devices that have had a huge impact on the study of novel optical, optoelectronic and optomechanical phenomena in the last couple of decades. One key targeted application has been that of all-optical modulation and control of light. Being solid state devices, however, their flexibility regarding their confinement characteristics and operating wavelength is limited, since it is typically defined by design and construction. In the most standard approach, highly confining resonators are used with the ultra-narrow cavity mode subtly tuned by small changes in refractive index \cite{Almeida,Preble}. A different situation would be attained if light trapping could not only be tuned but also set in an all-optical scheme. Such tunable 3D trapping of light is what we describe here.

\begin{figure*}[hhht]
\centering{}\includegraphics[width=140mm]{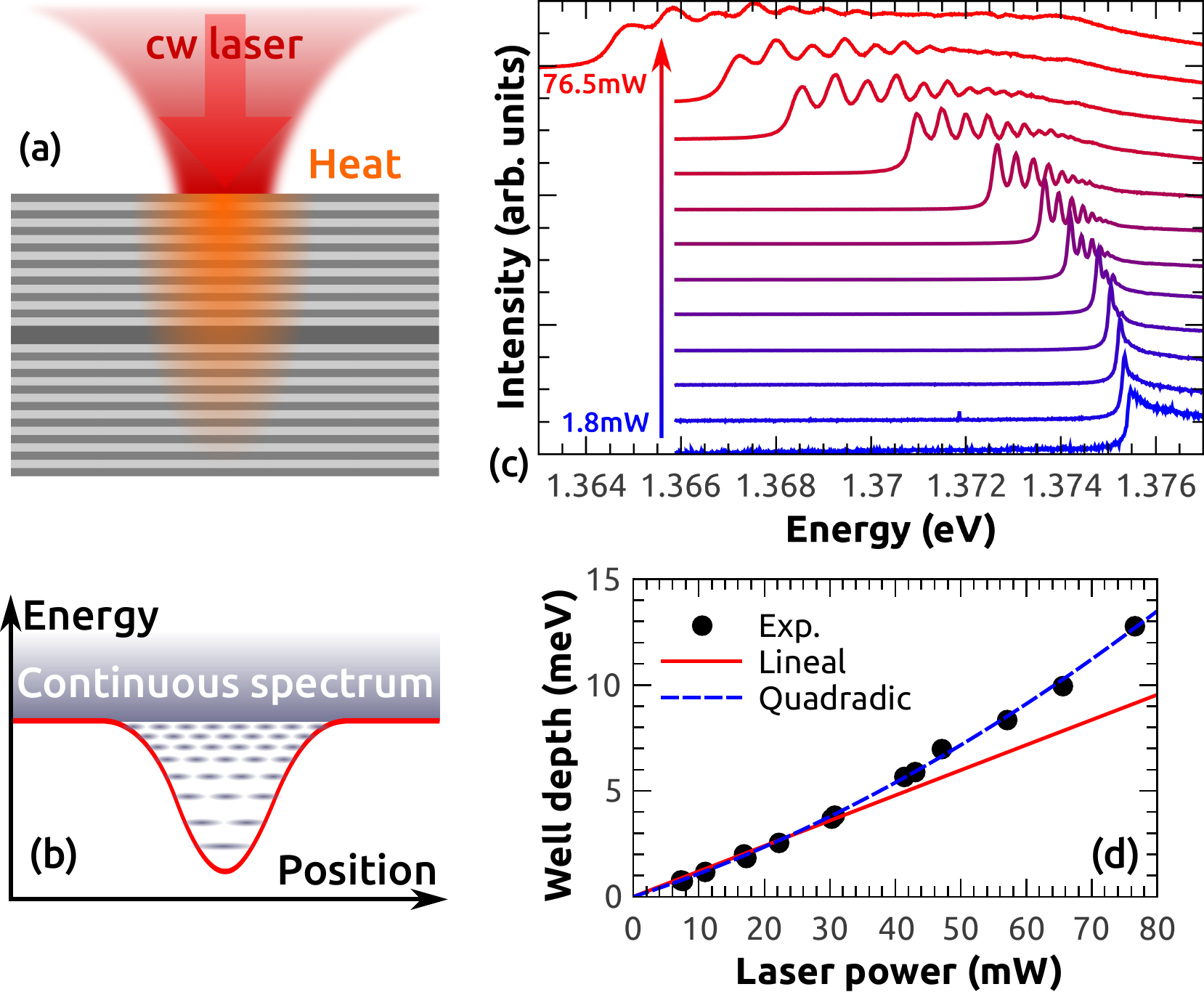}
\caption{a) Sketch of the system, showing the generation of electron-hole pairs and the resulting temperature distribution due to their relaxation with the lattice. b) Representation of a Gaussian optical well and the resulting confined modes, which appear beneath the original continuous spectrum. c) Photoluminescence spectra as a function of incident power, ranging from 1.8\mW \ up to 76.5\mW . d) Optical potential well depth as a function of laser power. The black circles are experimental results, while the red line is a linear fit to the low power ($\leq 30$\mW ) part of the data, and the blue dashed line is a quadratic fit to the whole data. \label{Fig1}}
\end{figure*}

Our approach to this task is based on a planar DBR semiconductor microcavity for the light confinement in one direction ($z$), plus a photon potential defined by a focused laser for the in-plane additional trapping (see the scheme in \figrefa{Fig1}). When a semiconductor is optically excited above or close to an electronic transition, a change of the refractive index occurs. This can be due to the modification of the electronic bands or the accessible electronic states induced by the presence of excited carriers \cite{Schmitt,Rink}, or by the lattice heating produced by the relaxation of these carriers \cite{CardonaGap}. If the change is {\em positive}, the local value of the cavity mode energy will be {\em decreased} and then light trapping will be possible. A conceptual scheme of the so-obtained 3D effective attractive photon potential is shown in \figrefb{Fig1}.

In the transparency region of a semiconductor, and for a fixed wavelength, the refractive index is expected to increase if the band gap is reduced. A gap reduction could be produced by an increase of the local temperature generated by photocarrier induced laser heating. However, more complex nonlinearities related to photoexcited carrier correlations could in principle also be involved \cite{Schmitt}. For example, polariton energy-barriers (in contrast with traps) have been demonstrated as a result of the blueshift of the polariton branch due to polariton-polariton interactions in microcavities at low temperatures and in the strong coupling regime \cite{BarrerasPolaritones,Wertz,Tosi}. Trapping can be induced, for example, by means of potential barriers generated with ring-like laser illumination, as reported for cold excitons in indirect quantum wells \cite{Hammack}.  Optically generated local traps in planar microcavities with attractive weak potentials have been shown to stabilize polariton condensates with a mechanism based on spatially confined gain \cite{PozosExcitones}, similar to gain guiding in semiconductor laser structures \cite{Salin}. Also, positive changes of the refractive index have been reported to affect photon lasing in GaAs microcavities when evolving from the excitonic regime toward the electron-hole plasma regime with increasing photoexcitation power \cite{Bajoni}. In contrast with these low-temperature carrier-related correlation effects, we will demonstrate that strong three dimensional trapping of photons can be achieved in planar semiconductor microcavities at room temperature through photothermal phenomena.

The studied sample consists of a high Q-factor $\lambda$/2 bulk-GaAs planar cavity enclosed by two DBRs consisting of alternating $\xtx{Ga}_{0.9} \xtx{Al}_{0.1} \xtx{As} / \xtx{Ga}_{0.05} \xtx{Al}_{0.95} \xtx{As}$ $\lambda$/4 layers, 28 pairs on the bottom, and 24 on top. The structure was grown by molecular beam epitaxy on a \simi 400\mum \ thick GaAs substrate. The structure presents a gradient in the layers' thicknesses that allows to tune the optical mode around the GaAs electronic transition at \simi 1.42\eV \ just by shifting the position of the laser spot on the sample. The reported experiments correspond to room-temperature photoluminescence microscopy measurements, with the optical cavity mode tuned below the GaAs gap. In high purity devices like the ones studied in this work the emission at energies below the electronic resonance is known to be mostly due to phonon assisted recombination \cite{Kurik,Khurgin2008,Khurgin2014}.

\begin{figure*}[hhht]
\centering{}\includegraphics[width=120mm]{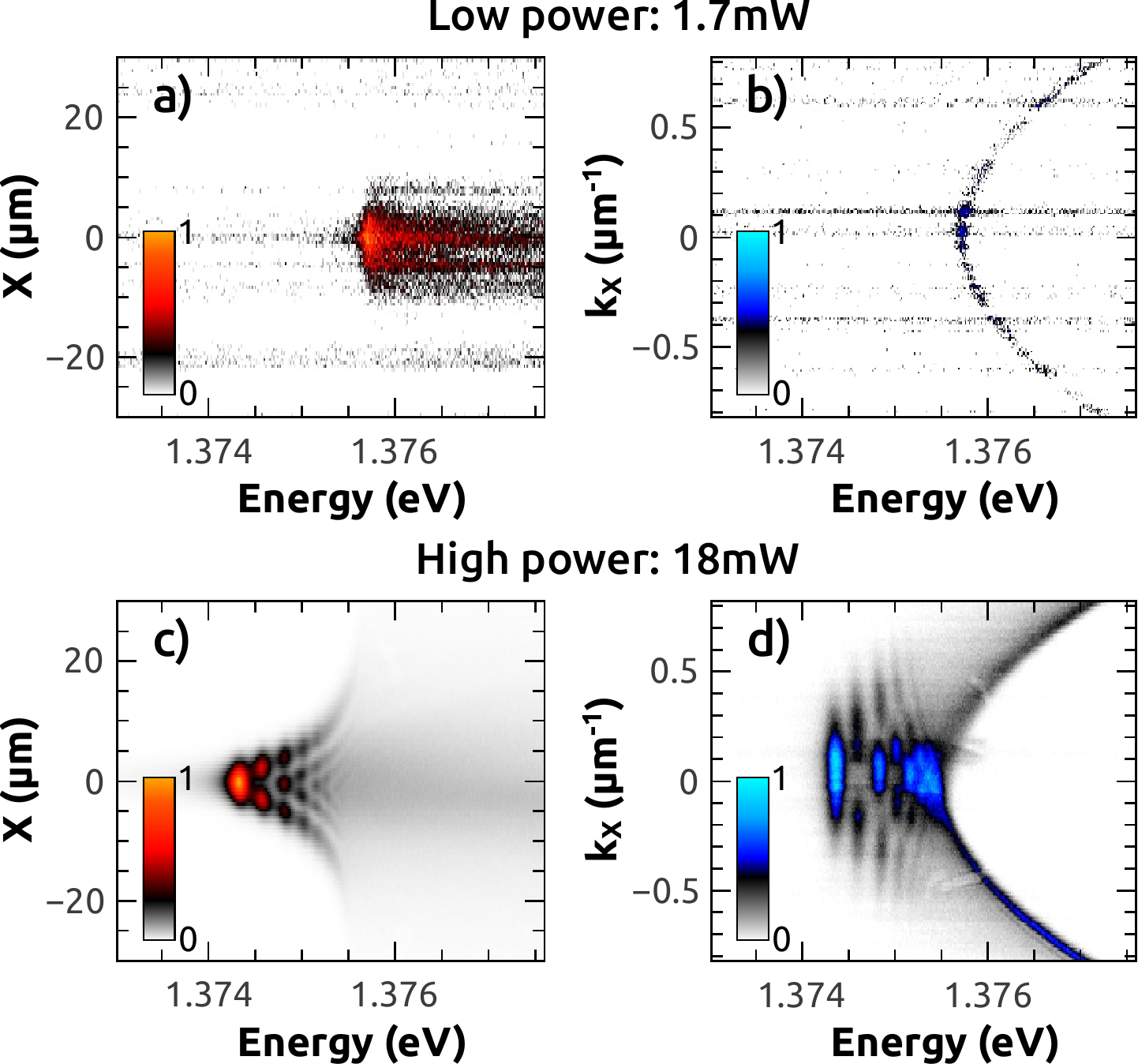}
\caption{Photoluminescence spectra spatial (a, c) and angular (b, d) resolution. The colors chosen for the spatial (orange) and the angular (blue) images are different to help on their rapid identification. The sample was excited with \simi 1.7\mW \ (a, b) and \simi 18\mW \ (c, d). \label{Fig2}}
\end{figure*}

The optical setup, described in the Supplemental Material\cite{SI}, allows for the acquisition of the integrated emitted spectra, or alternatively  the spatial or k-dispersion profile of the emission \cite{Aichinger}. The structure was fixed with heat-conducting silver paint on a copper holder in equilibrium with ambient temperature, and was excited with a 1.63\eV \ cw Ti-Sapphire laser. A 20x objective lens with a numerical aperture of 0.3 was used for both excitation and light collection, giving a Gaussian spot with full width at half maximum (FWHM) of \simi 10\mum . \figreffc{Fig1} presents a few examples of the photoluminescence spectra obtained when exciting with different incident powers, with the optical cavity mode tuned below but not far from the GaAs gap ($\varepsilon_{gap} \sim 1.42$\eV \ at room temperature), and collected over the full angular range of the objective (equivalent to $\xtx{k}_{r} \approx 2.1$\mum$^{-1}$ ).

\begin{figure*}[hhht]
\centering{}\includegraphics[width=150mm]{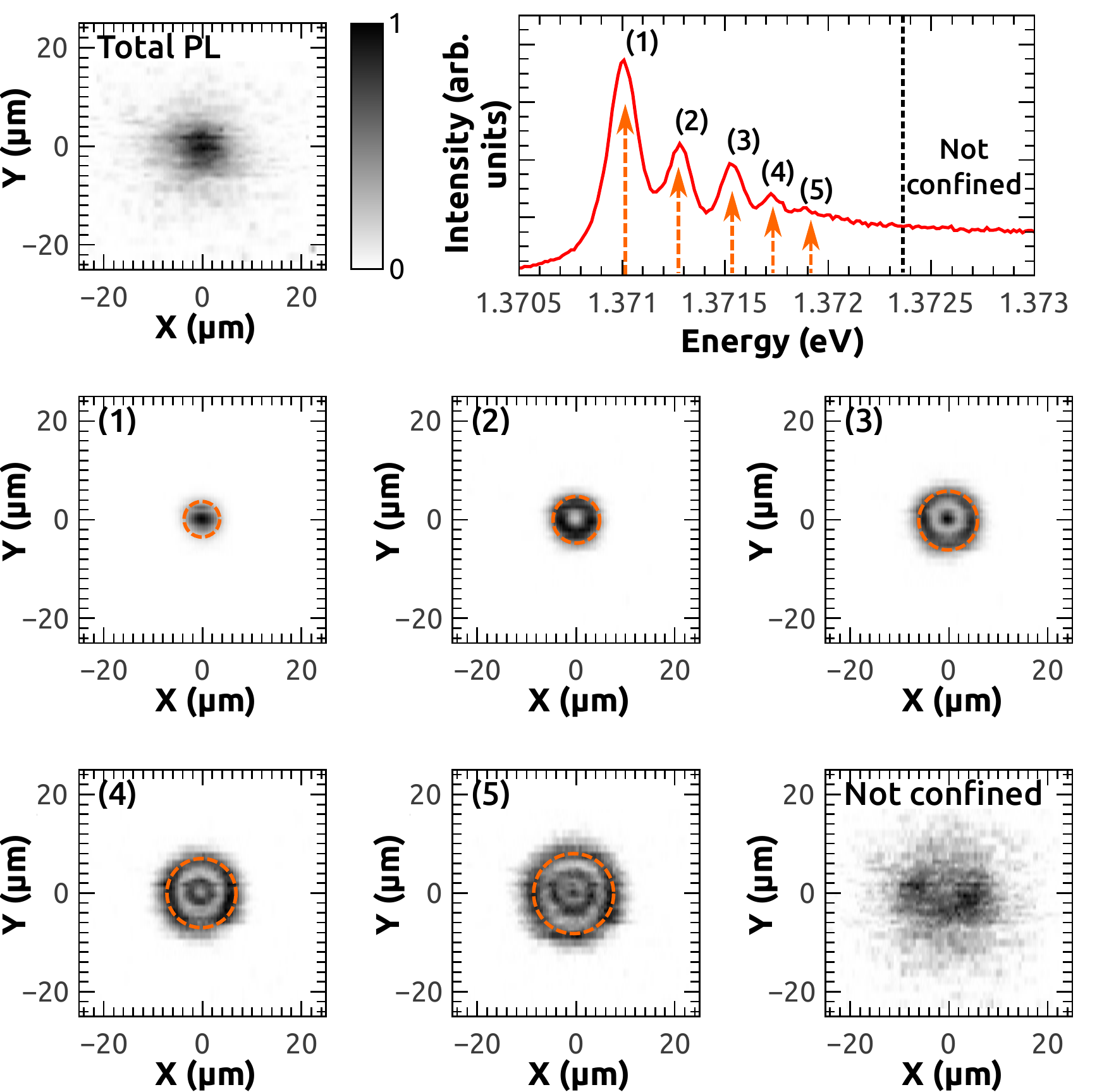}
\caption{Spatial distribution of the photoluminescence spectra for an excitation of \simi 18\mW . The top leftmost graph corresponds to the reconstructed spatial image of the emitted photoluminescence. The red curve on the top rightmost graph is the spectral distribution of the emitted light coming from the central line ($\xtx{Y} = 0$\mum) of the spot. The rest are different energy cuts (with incresing energy from left to right and top to bottom, labeled from 1 to 5). The dashed orange arrows and circles show the calculated energies and effective diameters of the confined modes, respectively. The vertical dashed line in the top-rightmost panel corresponds to $\varepsilon_{\infty}$. \label{Fig3}}
\end{figure*}

The low power spectrum in  \figrefc{Fig1} (1.8\mW) is representative of the planar microcavity (1D confinement) situation. The planar cavity quality factor in the transparency region, below the GaAs gap, is around $1 \times 10^4$. This corresponds, for the emission normal to the structure, to a narrow peak of \simi 150\mueV \ width centered in this case at \simi 1.3755\eV . The confined cavity mode emission along $z$ defines the low energy flank of this curve. Its flat continuation towards higher energies is determined by the in-plane parabolic dispersion of the planar cavity optical mode, which is collected within the numerical aperture of the microscope objective \cite{Houdre}. Major and quite striking changes in the spectra occur when the focused laser power is increased. Some conspicuous new features arise, namely  a red shift of the emission, and the appearance of clear peaks contrasting with the essentially featureless low-power spectrum. As we argue next, these features  evidence the optical confinement within a Gaussian photon potential well, which is induced by the local heating of the sample due to the laser excitation. The trapping energy can attain quite significant values under the focusing conditions used, e.~g. \simi 13\meV \ for a \simi 76.5\mW \ excitation, a value equivalent to 100 times the FWHM of the unperturbed planar cavity mode.

Three dimensional confinement of light should be reflected in the induced changes of the spectra (as manifested by the energy lowering and spectral discretization in \figrefc{Fig1}), but also by the corresponding modified spatial and k-space (angular) characteristics of the photoluminescence emission \cite{KspaceDisc}.  \figreff{Fig2} presents the spatial (a, c) and angular (b, d) distributions of the emitted light when a low (a, b) or high (c, d) power excitation is applied. Again, the low power panels (1.7\mW) are representative of the planar (1D confinement) case. As expected, the spatial distribution reproduces the shape of the excitation Gaussian profile. Its in-plane k-dispersion in turn reflects the typical parabolic behavior of light confined in one dimension, but free to propagate in the other two (in-plane) directions. The situation dramatically changes when increasing the focused laser power (18\mW \ in \figref{Fig2}). New discretized modes appear at lower energies, with spatial and in-plane k-dispersion strongly resemblant of laterally confined modes, as seen e.g. in micropillar cavities \cite{Pilares,SizeDependence,KspaceSquare} or in laterally microstructured planar resonators \cite{Tamm2,Paulo}. The spatial and spectral envelope of the observed modes in \figrefc{Fig2} nicely reflects the Gaussian photon potential trap induced by the focused laser.

\begin{figure}[hhht]
\centering{}\includegraphics[width=90mm]{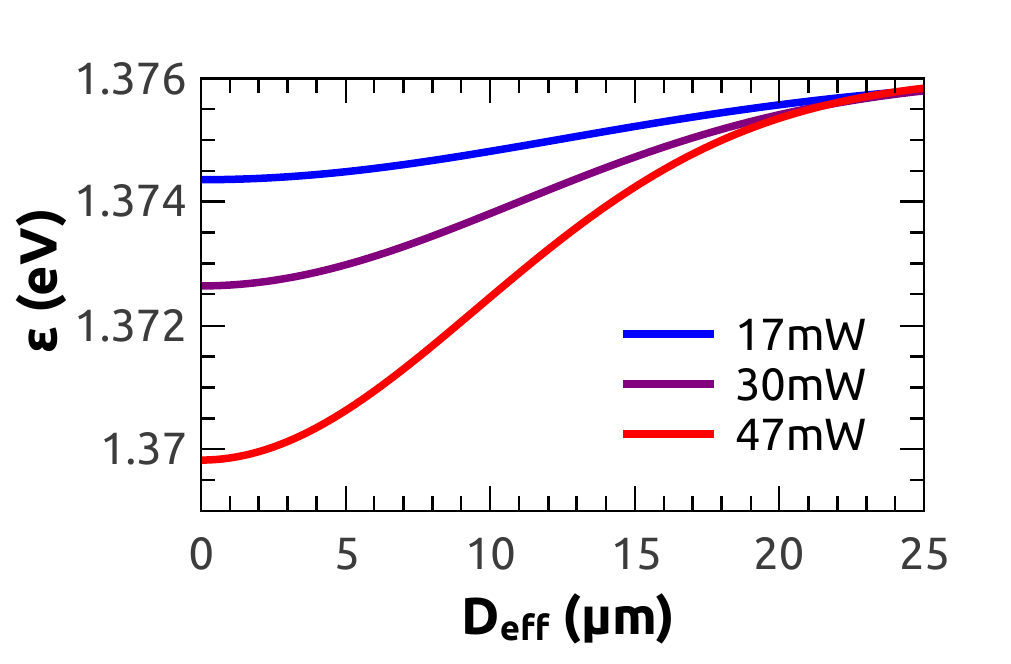}
\caption{Comparison between the obtained optical potential wells for three different excitation powers. \label{f:PozoP}}  
\end{figure}

\begin{figure*}[hhht]
\centering{}\includegraphics[width=120mm]{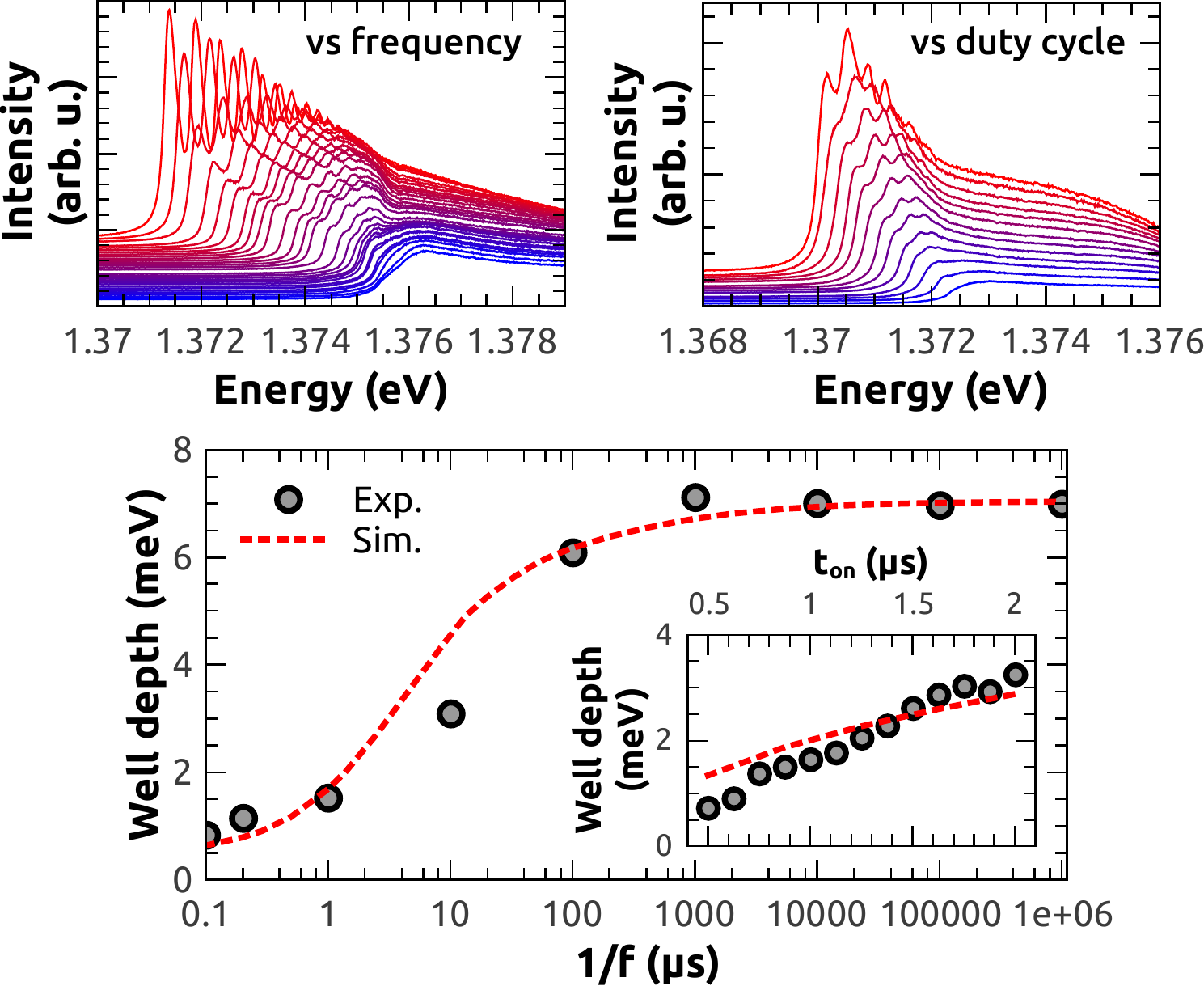}
\caption{Top panel: vs frequency) photoluminescence spectra as a function of frequency for a 20{\%} duty cycle square-wave modulation, with excitation power of $41.3$\mW . From blue to red, the curves correspond to frequencies ranging from 1\MHz \ to 1\Hz . Note that the selected frequencies are not equidistant; e.~g. the three most shifted spectra correspond to 1\Hz , 1\kHz \ and 5\kHz . vs duty cycle) photoluminescence spectra as a function of the duty cycle for a square-wave modulation of 400\kHz \ and an excitation power of 33\mW , taken at a slightly different position on the sample. From blue to red, the curves correspond to duty cycles ranging from 20{\%} to 80{\%}, with steps of 5{\%}. 
Bottom panel: main frame) optical well depth as a function of the modulation period for an excitation of 47\mW \ and a duty cycle of 20{\%}, taken at a slightly different position on the sample. Inset) optical well depth as a function of the duty cycle time. The circles correspond to the experimental results, taken from the curves shown in the top rightmost panel, while the red dotted curves were obtained from a thermal diffusion model.\label{Fig4}}
\end{figure*}

In \figref{Fig3} we show the lateral distribution of the emitted light for 18\mW \ excitation. The top left graph corresponds to the total integrated emission, reflecting the homogeneous spatial distribution of the excitation beam. The spectrum displayed in the top-right panel corresponds to this emission when the spot is imaged centered on the spectrometer entrance slit. The vertical dotted line in this latter panel shows the energy of the unperturbed cavity mode, above which modes are not confined. The discrete peaks (1)-(5) correspond to the observed fully 3D confined photonic states. The six remaining graphs in this figure are constant energy sections that were selected to leave only one mode for each cut (labeled with the same numbers 1 to 5), and the non-confined spectral region. Circularly symmetric modes confined within different radial distances are clearly identified.

The spectral and spatial distributions of the observed modes have the required information to model the laser-induced photon potential. Regardless of the exact physical process, which we will approach later on, it is apparent that the laser excitation is generating a local change in the optical properties of the structure, which in turn produces an optical potential well that confines light in an analogous way as that observed in a micropillar structure \cite{Pilares}. Based on the excitation profile and symmetry of the studied system, a Gaussian well of finite depth is expected to be generated:
\begin{equation}
\varepsilon = \varepsilon_{\infty}-\Delta \varepsilon_0 e^{-\frac{\xtx{r}^2}{2 \sigma^2}},
\label{ec:4.18bis}
\end{equation}
where r is the radius, $\sigma$ the standard deviation $\varepsilon_{\infty}$ the energy of the continuum (datum obtained from the experiment), and $\Delta \varepsilon_0$ the effective depth of the well. The problem can be significantly simplified by making the approximation that each confined mode of energy $\varepsilon$, interacts with an infinitely deep circular well of diameter $\xtx{D}_{eff}$, given by \eqref{ec:4.18bis}. This is equivalent to considering that the confined modes present a node on the wall of the Gaussian well, if we define $\xtx{D}_{eff}$ so that it follows the Gaussian shape of the actual potential well. Therefore, reformulating \ecref{ec:4.18bis}, we get the dependence between the mode energy and the effective diameter of the infinitely deep well that best approximates the problem:
\begin{equation}
\varepsilon = \varepsilon_{\infty}-\Delta \varepsilon_0 e^{-\frac{\xtx{D}_{eff}^2}{8 \sigma^2}},
\label{ec:4.18}
\end{equation}
being $\Delta \varepsilon_0 = \varepsilon_{\infty}-\varepsilon_0$ the potential well depth and $\varepsilon_0$ the energy of the bottom of the potential well.

Under these assumptions, the confined modes can be well represented with Bessel functions, and their energies can be calculated as \cite{KspaceDisc}:
\begin{equation}
\varepsilon = \sqrt{{\varepsilon_0}^2 + \frac{4 \hbar^2 {\xtx{c}}^2}{{\xtx{n}_{cav}}^2} {\frac{\xtx{X}_{ml}^2 }{\xtx{D}_{eff}^2}}  },
\label{ec:4.17}
\end{equation}
where $\xtx{X}_{ml}$ is the m-th root of the Bessel function of order l, c the speed of light in vacuum and $\xtx{n}_{cav}$ the effective index of refraction of the sample. \ecreff{ec:4.18} and \ecref{ec:4.17} can be combined to obtain $\varepsilon_0$ and $\sigma$ as fitting parameters, with the measured energies $\varepsilon$ and the corresponding Bessel roots $\xtx{X}_{ml}$ as the input data (see the Supplemental Material\cite{SI} for a more detailed description of the procedure). The theoretical results reproduce simultaneously the effective diameters and the peaks of the observed spectra very well (see for example the dashed circles and arrows in \figref{Fig3}).

In \figref{f:PozoP} the effective photon potentials obtained for three different excitation powers are compared. A deepening of the well is observed as the power is increased, while the width turns out to be fairly constant, as expected from the invariant laser spot shape. This agreement allows to conclude that, at least for low excitation powers, the laser-induced potential shape is essentially independent of the laser power (its Gaussian shape and lateral width $\sigma$ remain unaltered), being only the well depth strongly sensitive to the excitation power. In fact, as can be observed in \figrefd{Fig1}, the photon potential depth shows a supra linear dependence with laser power above \simi 40\mW . In principle the number of photoexcited carriers should be proportional to the incident power. The observed quadratic dependence might be reflecting an increased absorption at the fixed laser energy, e.g. due to a GaAs gap reduction induced by a temperature increase.

This latter result points towards the physical origin of the proposed Gaussian photon potential. The potential, schematized in  \figrefb{Fig1}, represents the effective local energy of the optical cavity mode of an equivalent planar structure. It is this lateral gradient of the cavity mode energy that leads to light confinement and to new photonic modes (just like a quantum well leads to new confined electronic states defined by, though shifted from, the confinement potential). Similar ideas have been applied to model effective phonon potentials \cite{Balseiro}. The resulting modes are red-shifted (with respect to the unperturbed planar structure) by the laser excitation, mapping out the shape of the focused laser intensity profile (see for example \figrefc{Fig2}). As mentioned in the introduction, either electronic or thermal phenomena can be at the basis of the observed phenomena. The different phenomena have however very different dynamics, typically reflecting pico- and nanosecond recombination when photoexcited carriers are responsible \cite{PRAelectrons}, and microsecond thermal diffusion times if heating is dominant \cite{Thermal}. To clarify this issue, the dynamics of the photon potential was studied using modulated optical excitation. The modulation was achieved with an acousto-optic modulator, fed with a square wave generator of tunable frequency and duty cycle.

The top left panel in \figref{Fig4} presents photoluminescence spectra as a function of the modulation frequency, for a duty cycle of $20 \%$. From blue to red, the curves correspond to modulation frequencies ranging from 1\MHz \ to 1\Hz . The figure clearly shows the 3D confined spectra for slow modulation (similar to the cw experiments shown above), which evolve into the laterally unconfined planar situation with increasing modulation frequency. The top right panel in \figref{Fig4} (corresponding to a slightly different position on the sample) displays the photoluminescence spectra variation with duty cycle for a square wave modulation of 400\kHz . This particular frequency was selected because it is where the discrete modes start to be noticed for the minimum duty cycle available (20\%). From blue to red, the curves correspond to duty cycles ranging from 20{\%} (0.5\mus \ on, 2.0\mus \ off) to 80{\%} (2.0\mus \ on, 0.5\mus \ off), with steps of 5{\%}. The times involved in the observed dynamics clearly point towards a thermal phenomenon. We note that these results can be related to the ``thermal lensing'' previously investigated in vertical cavity surface emitting lasers (VCSELs), by which the laser mode lateral distribution is changed depending on the drive power, due to non-uniform heating \cite{Brunner,4628733}. 

Having established the thermal origin of the phenomena, we note that two different effects could in principle be responsible of the observed features: a local increase of the refractive index, or a local expansion of the structure. Calculating the spectral position of the optical mode with the transfer matrix method (considering the expansion coefficient and the dependence on temperature of the refractive index of the materials), it turns out that, for a given increase in temperature, the effect of the refractive index change on the spectral position of the optical mode is roughly \simi $8$ times larger than that due to the thermal expansion of the lattice. 

Assuming that the change in the refractive index that gives rise to the optical well is directly proportional to the local change in temperature due to the laser excitation, it is possible to describe the physical situation as a heat diffusion problem. A thermal diffusion model was used to obtain the structure's mean temperature during excitation, as a function of the modulation frequency and duty cycle (see the Supplemental Material\cite{SI} for a detailed description of the calculations). The thermal diffusivities used in the model were taken from the literature, and are included in the Supplemental Material\cite{SI}. Since the exact fraction of the incident laser power that is actually converted into heat is difficult to determine, we used the thermal source maximum value (at the top $\xtx{Ga}_{0.9} \xtx{Al}_{0.1} \xtx{As}$ layer) as the (only) fitting parameter. 

In the bottom panel in \figref{Fig4} we show a comparison between the experimentally observed optical potential well depth (circles), derived from  experimental curves as in the top left figure (though taken at a different power and position on the sample) and the theoretical results (red dashed line) obtained from the thermal diffusion model (maximum heating power of $1.0\frac{\xtx{mW}}{\mu \xtx{m}^3}$), as a function of the inverse of the modulation frequency f. The data in the inset corresponds to the well depth as a function of the duty cycle time, obtained from the measurements shown in the top right panel, again compared with the theoretical curve derived from the thermal diffusion model (maximum heating power of $0.5\frac{\xtx{mW}}{\mu \xtx{m}^3}$). In order to estimate the optical potential depth for these measurements, there are two possibilities. If the confined modes are well defined in energy, the method explained before (Bessel modes) could be used; if the modes are somewhat blurred, as occurs for high frequency modulation, or are too weakly confined to appreciate them, it is still possible to estimate the potential well depth from the envelope function of the emitted photoluminescence. The two methods give similar results, but to be able to compare the high and low frequency modulations in \figref{Fig4}, as well as the long and short duty cycles, the second one was used. The agreement between experiment and theory shown in \figref{Fig4} clearly confirms thermal effects at the origin of the reported light trapping phenomena. 

To finish, we note that equivalent 3D confinement was obtained using a 514\nm \ laser, which is strongly absorbed at the first layers of the top DBR, thus only weakly coupling with the rest of the structure. This implies that two-laser schemes may be implemented, with one laser used to seed the cavity emission, and the other to dynamically control the 3D light trapping. We note that the GaAs gap strongly depends on temperature above \simi 100\K \cite{SSC}. Below this temperature electronic nonlinearities (which are typically weaker but faster) should be dominant over the observed thermal effects. With both approaches rich photon potentials could be accessible using phase and intensity modulated illumination as that used in super-resolution microscopy and other photonic applications.

{\bf Acknowledgments:} This work was partially supported by the ANPCyT Grants PICT 2012-1661 and 2013-2047, the French RENATECH network, and the international Franco-Argentinean Laboratory LIFAN (CNRS-CONICET).

{\bf Author information:} Correspondence and requests for materials should be addressed to afains@cab.cnea.gov.ar


%


\onecolumngrid
\newpage
\setcounter{figure}{0}  
\renewcommand{\figurename}{FIG. \!}
\renewcommand{\thefigure}{S\arabic{figure}}


\begin{center}
\underline{\huge{\textbf{Additional Information}}}
\end{center}
\vspace{0.5cm}

In this supplemental material we provide a brief description of the experimental setup, and details on the theoretical models used to explain the spatial and spectral distribution of the optically confined modes, as well as their thermal origin. 

\section{Experimental setup}

{\begin{figure}[hhht]
\centering{}\includegraphics[width=0.49\textwidth]{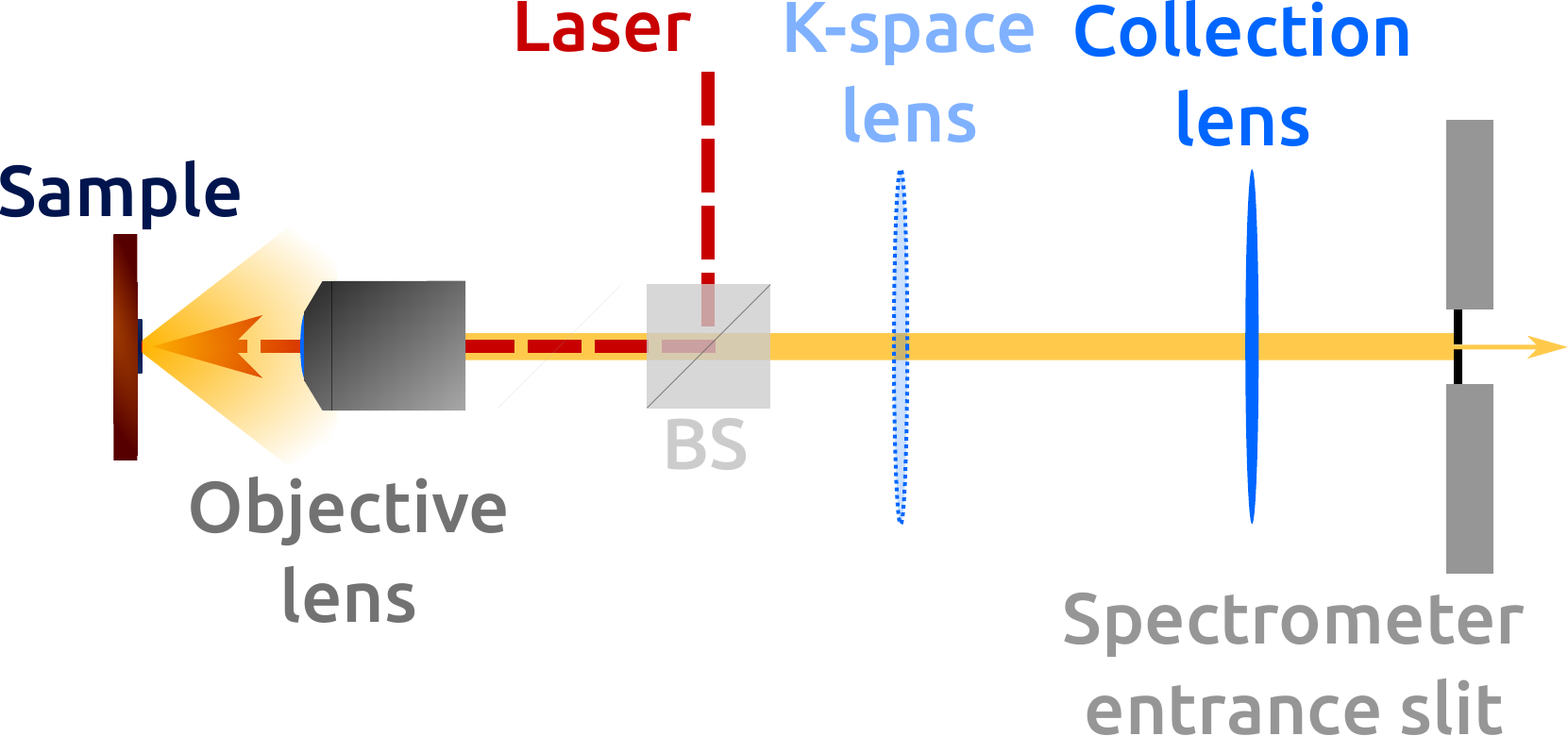}
\caption{Experimental setup. The laser, represented with a red dashed line, is reflected by a beam-splitter and focused on the sample by a microscope objective. The emitted photoluminescence (yellow) is partially collected by the same objective lens and imaged on the spectrometer entrance slit by the collection lens. An additional lens can be used to image the k-space distribution of the photoluminescence. \label{f:setup}}  
\end{figure}}

The experimental setup used, outlined in \figref{f:setup}, consists of a 20x and 0.3NA microscope objective, a beam splitter and a set of two lenses: the collection lens, and an optional k-space-imaging lens. A Spectra Physics Ti-Sapphire continuous wave laser is used for the excitation (red dashed line in \figref{f:setup}). The laser energy was set to 1.63\eV . The emitted photoluminescence (yellow) is collected by the same objective and imaged on the spectrometer entrance slit by the collection lens. The latter is movable in the direction perpendicular to this slit, enabling the acquisition of the spectra coming from different spatial positions.

\FloatBarrier

\begin{figure*}[hhht]
\centering{}\includegraphics[width=140mm]{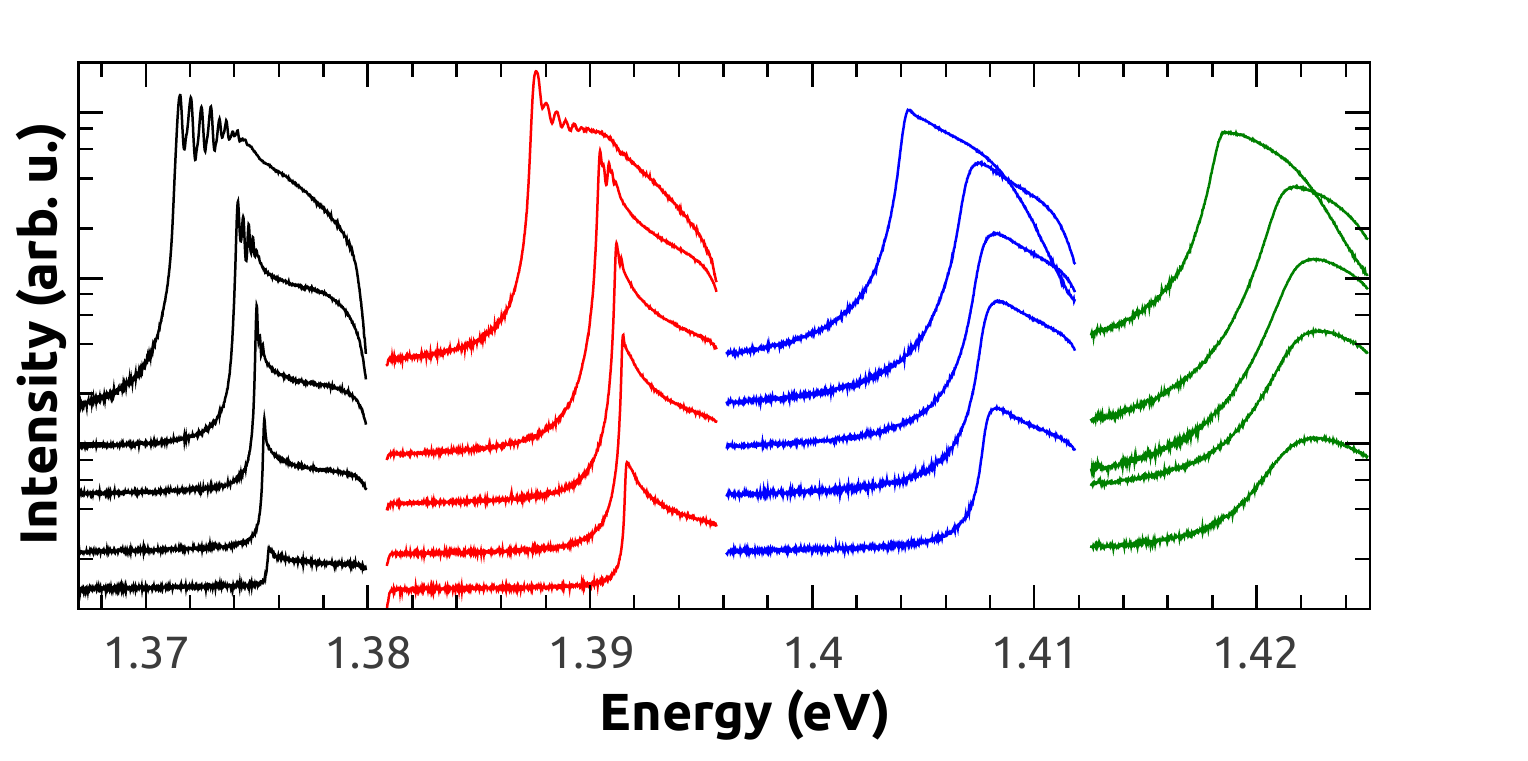}
\caption{Photoluminescence spectra obtained for 4 different spectral positions of the optical mode, corresponding to different positions on the sample. The curves of the same color correspond to same optical mode tuning. The five curves for each tuning position correspond to different excitation powers (1.5\mW , 4.5\mW , 9\mW , 18\mW \  and 40\mW).
\label{f:ABCD}}  
\end{figure*}

The optional k-space lens can be positioned so that its focus coincides with the microscope objective back-focus. This technique, usually referred to as ``k-space imaging'', is based on the same ideas as that of reciprocal space imaging in transmission electron microscopy \cite{TEM}. Instead of imaging the real image of the sample, the collection lens images the plane corresponding to the microscope objective back-focus. All the light coming from the sample at a certain angle is focused on the same point on this plane. Therefore, the image formed at the spectrometer entrance slit has the different angular components of the emitted photoluminescence spatially separated, enabling the spectral analysis of the emission's dispersion relation.

Photoluminescence measurements with different excitation powers were carried out for 4 different spectral positions of the cavity optical mode with respect to the electronic resonance of GaAs (\simi 1.42\eV). In \figref{f:ABCD} the obtained spectra are presented. The green curves correspond to the case in which the cavity mode is tuned on the resonance, and therefore the material presents maximum absorption, while the black curves correspond to a tuning such that the sample is essentially transparent. Several differences are observed between these spectra. At low excitation power there is a significant difference in the intensity of the emission, larger with increasing energy, as expected from photoluminescence near an electronic resonance \cite{Cardona2}. At high powers there is no marked difference in the intensity of the emission, but a clear difference in the spectra is observed. In the case that the mode is tuned in the transparency zone (black curves), the confined modes are clearly visible. As the optical mode is tuned to the resonance, the widths of the confined modes begin to increase, until they become completely indistinguishable (green curves). In this case, the resulting spectra present the expected appearance of an emission dominated by the heating of the sample, where the optical mode suffers a decrease in its quality factor and a shift towards lower energies. 

\section{Optical potential well model}

In the main text we presented the relation between the energy of the confined modes and the effective diameter of the infinitely deep potential well, which we assume follows the profile of a Gaussian potential generated by the laser excitation:
\begin{equation}
\varepsilon = \varepsilon_{\infty}- (\varepsilon_{\infty}-\varepsilon_0) e^{-\frac{\xtx{D}_{eff}^2}{8 \sigma^2}},
\label{ec:4.18}
\end{equation}
as well as the energy dependence of the confined Bessel modes:
\begin{equation}
\varepsilon = \sqrt{{\varepsilon_0}^2 + \frac{4 \hbar^2 {\xtx{c}}^2}{{\xtx{n}_{cav}}^2} {\frac{\xtx{X}_{ml}^2 }{\xtx{D}_{eff}^2}}  }.
\label{ec:4.17}
\end{equation}
We use \ecref{ec:4.18} and \ecref{ec:4.17} to calculate the energy of the photonic modes trapped in the Gaussian effective potential. For reasons not yet understood, not all Bessel modes are observed experimentally. As in previous work \cite{PozosExcitones}, we used the Bessel modes that best described the spatial distribution of the experimentally observed modes.

\begin{figure}[hhht]
\centering{}\includegraphics[width=0.49\textwidth]{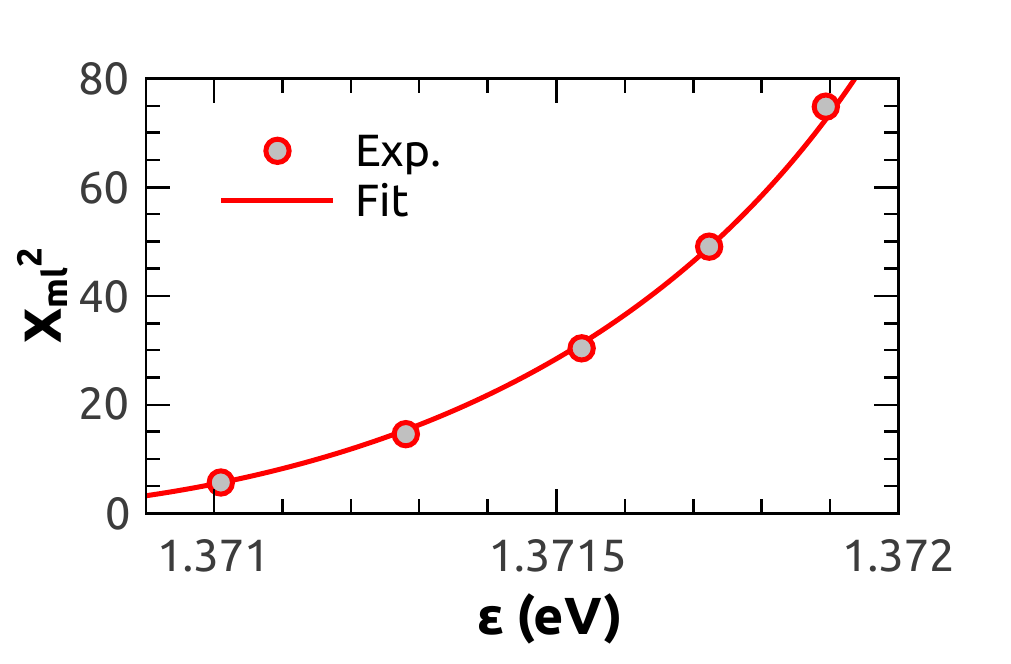}
\caption{Example of the experimental points and the fitting curve obtained with \eqref{ec:4.17b}. These data correspond to the results shown in Fig. 3 of the main text. \label{f:fiteo}}  
\end{figure}

We proceed as follows. $\xtx{D}_{eff}$ can be eliminated from \ecref{ec:4.18} and \ecref{ec:4.17} to obtain:
\begin{equation}
\xtx{X}_{ml}^2 = 2 \frac{{\xtx{n}_{cav}}^2}{ \hbar^2 {\xtx{c}}^2} \sigma^2 (\varepsilon^2-\varepsilon_0^2) \xtx{ln}\left( \frac{\varepsilon_{\infty}-\varepsilon_0}{\varepsilon_{\infty}-\varepsilon} \right).
\label{ec:4.17b}
\end{equation}
Since we experimentally know $\varepsilon$ for each mode (characterized by a specific Bessel order), \ecref{ec:4.17b} can be used to fit the data, using $\sigma$ and $\varepsilon_0$ as fitting parameters. An example is presented in \figref{f:fiteo}.

\FloatBarrier

\begin{figure}[hhht]
\centering{}\includegraphics[width=0.49\textwidth]{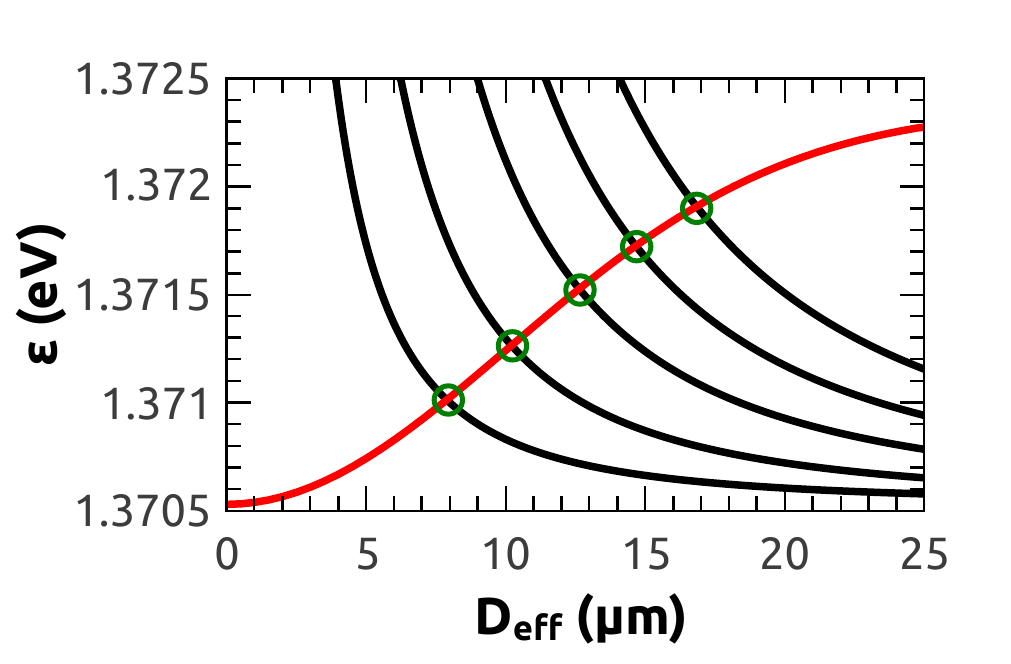}
\caption{Confinement dependence of Bessel type modes (black) and effective optical potential well (red). The solutions are given by the crossings between the curves, highlighted with the green hollow circles. These theoretical values are the ones presented in Fig.~{\color{azul}3} of the main text. \label{f:teoria}}  
\end{figure}

Having determined $\sigma$ and $\varepsilon_0$, \ecref{ec:4.18} and \ecref{ec:4.17} can be used to obtain the theoretically determined values of the energy $\varepsilon$ and effective diameter $\xtx{D}_{eff}$ of the confined modes. This is easiest done through a graphical analysis, where the solutions to the problem, given by the values of $\varepsilon$ and $\xtx{D}_{eff}$ for each mode, are given by the crossings between the curves corresponding to \ecref{ec:4.18} and \ecref{ec:4.17}. An example is presented in \figref{f:teoria}. These so determined values of $\varepsilon$ and $\xtx{D}_{eff}$ are the ones shown in Fig.~{\color{azul}3} of the main text with dashed arrows and circles, respectively.

\section{Thermal diffusion model}

{\begin{figure*}[hhht]
\centering{}\includegraphics[width=140mm]{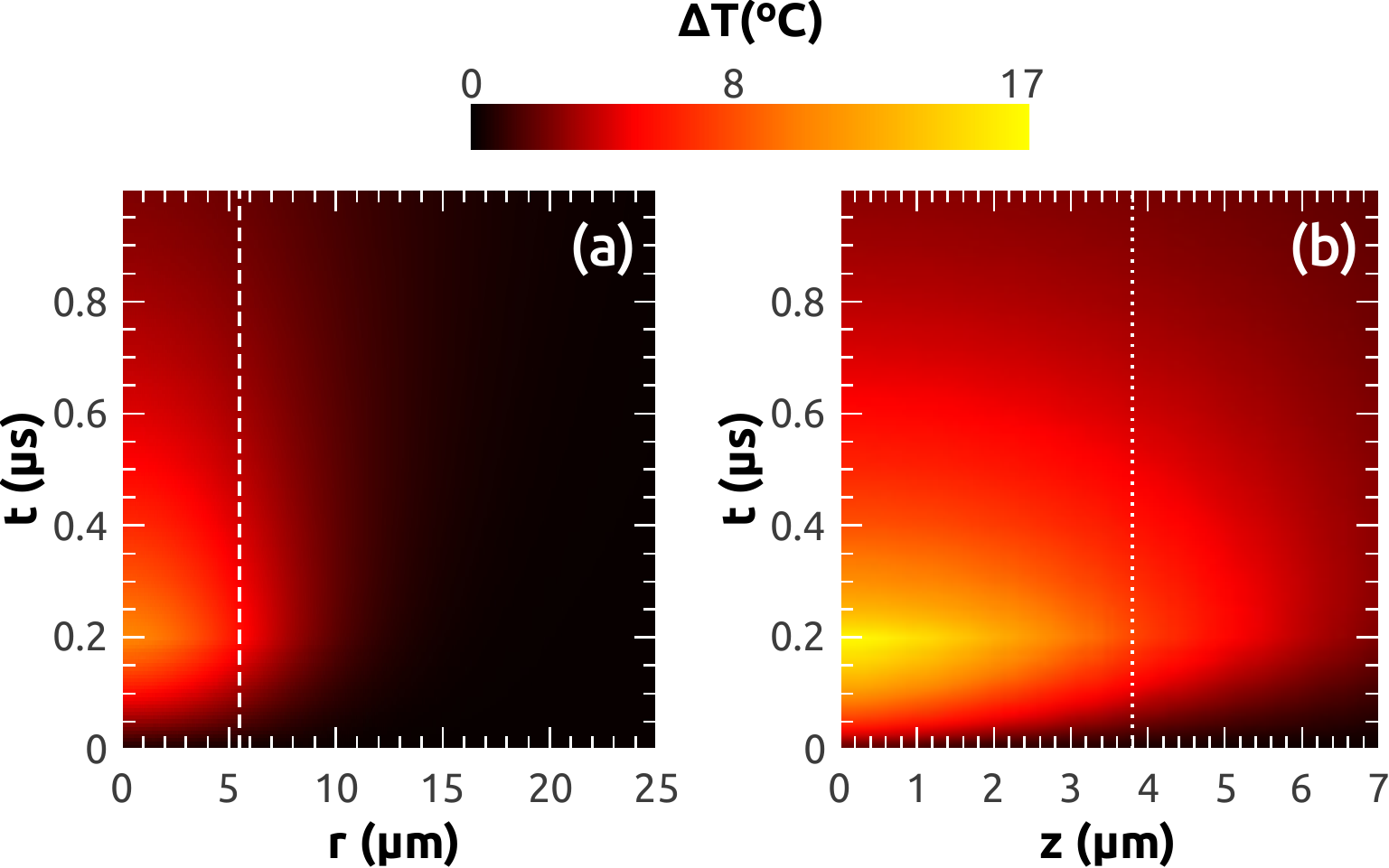}
\caption{Finite elements simulation (COMSOL) for the heat diffusion problem. a) Radial temperature distribution at the GaAs spacer ($\xtx{z} = \xtx{z}_\xtx{cav}$) as a function of time for a Gaussian heat source with $\xtx{FWHM} \approx 10$\mum \ and with an exponential decay dependence in the z direction. The white dashed line marks the position of the half-maximum of the heat source. b) Temperature distribution in the growth direction at the center of the heat source (r = 0). The white dotted line marks the position of the GaAs spacer. The source is turned on at $\xtx{t} = 0$\mus \ and off at $\xtx{t} = 0.2$\mus . This simulation corresponds to a modulation frequency of 1\MHz \ and a duty cycle of 20\%. \label{f:Maps}}  
\end{figure*}}

Assuming that the change in the refractive index that gives rise to the optical effective potential is directly proportional to a local change in temperature due to the laser excitation, it is possible to describe the system as a heat diffusion problem \cite{DifT2}:
\begin{equation}
\frac{\partial \xtx{T(\xtx{x},\xtx{y},\xtx{z},t)}}{\partial \xtx{t}} = \alpha_T \nabla^2 \xtx{T(\xtx{x},\xtx{y},\xtx{z},t)} + \xtx{Q(\xtx{x},\xtx{y},\xtx{z},t)},
\label{e:dif2}
\end{equation}
being T the temperature, $\alpha_T$ the thermal diffusivity and Q a source term. Several approximations can be performed to provide a clear picture of the physics involved. Let us consider first the source term. The temperature change is due to the relaxation of photoexcited carriers. Therefore, the temporal and spatial dependence of Q will be given by that of the e-h pairs excited by the incident laser. Because of the magnitude of the electronic gap, close to the laser excitation, the GaAs and the $\xtx{Ga}_{0.9} \xtx{Al}_{0.1} \xtx{As}$ layers are the only ones that will actually be excited by the 1.63\eV \ laser. Furthermore, the Al-rich layers on top and below them act as energy barriers (because of the higher gap), so carriers will be confined inside each layer until they recombine. The heat source can therefore be considered not null only on the absorbing layers. In the lateral direction, the distribution will be given by the convolution of the excitation profile and the diffusion of the photoexcited carriers (i.~e a Gaussian distribution with a $\xtx{FWHM} \approx 10$\mum \ in our case \cite{PRAelectrons}).

Since the characteristic recombination times for photoexcited e-h pairs in these structures are around 7\ns \ \cite{PRAelectrons}, and the modulation times we are working with are of the order of the \mus , the heat generation can be considered instantaneous. That is, a photon will be instantaneously converted into heat at the spacer, and thus the time dependence of Q will closely follow the laser modulation.

The heat source $ \xtx{Q(x,y,z,t)}$ can then be approximated as:
\begin{widetext}
\begin{equation}
 \xtx{Q(x,y,z,t)} = \xtx{q(t)} \cdot e^{-\frac{\xtx{x}^2+\xtx{y}^2}{2 \sigma^2}}  \cdot \sum e^{-\xtx{z}_i/\xtx{z}_0} \bigg( \Theta \left( \xtx{z}_i \right) \Big( 1-\Theta \left( \xtx{z}_i+\xtx{d}_i \right) \Big) \bigg),
\label{e:dif3}
\end{equation}
\end{widetext}
where  $\xtx{q(t)}$ describes the temporal modulation of the excitation laser, $\sigma$ is the standard deviation of the in-plane Gaussian distribution and the last term describes the dependence in the growth direction. The latter is given by an exponential decay, due to photon absorption as the laser penetrates the sample (being $\xtx{z}_0$ the penetration depth); besides, there is only heat generation in the Ga-rich layers, as already mentioned, which is represented by the product between Heaviside functions $\Theta$, being $\xtx{z}_i$ and $\xtx{d}_i$ the position and thickness of these layers, respectively.

A finite-elements simulation for this heat diffusion problem was performed with the COMSOL software. A heat source with Gaussian lateral distribution and with maximum value of $0.74\frac{\xtx{mW}}{\mu \xtx{m}^3}$, $\xtx{FWHM} \approx 10$\mum \ and an exponential decay dependence on the growth direction was considered, which is only active in the Ga-rich layers, for the reasons mentioned before. The penetration depth $\xtx{z}_0 = 2.2$\mum \ was derived using a transfer matrix method to model the light propagation along the axis of the structure, using published values for the refractive index and absorption of the materials. The results are presented as two color maps in \figref{f:Maps}, showing the temperature distribution as a function of time in the radial (\figrefa{f:Maps}) and growth (\figrefb{f:Maps}) direction. It is clear that most of the heat diffuses towards the substrate, while the temperature profile in the radial direction changes mostly in amplitude, but not so much in shape. 

{\begin{figure*}[hhht]
\centering{}\includegraphics[width=120mm]{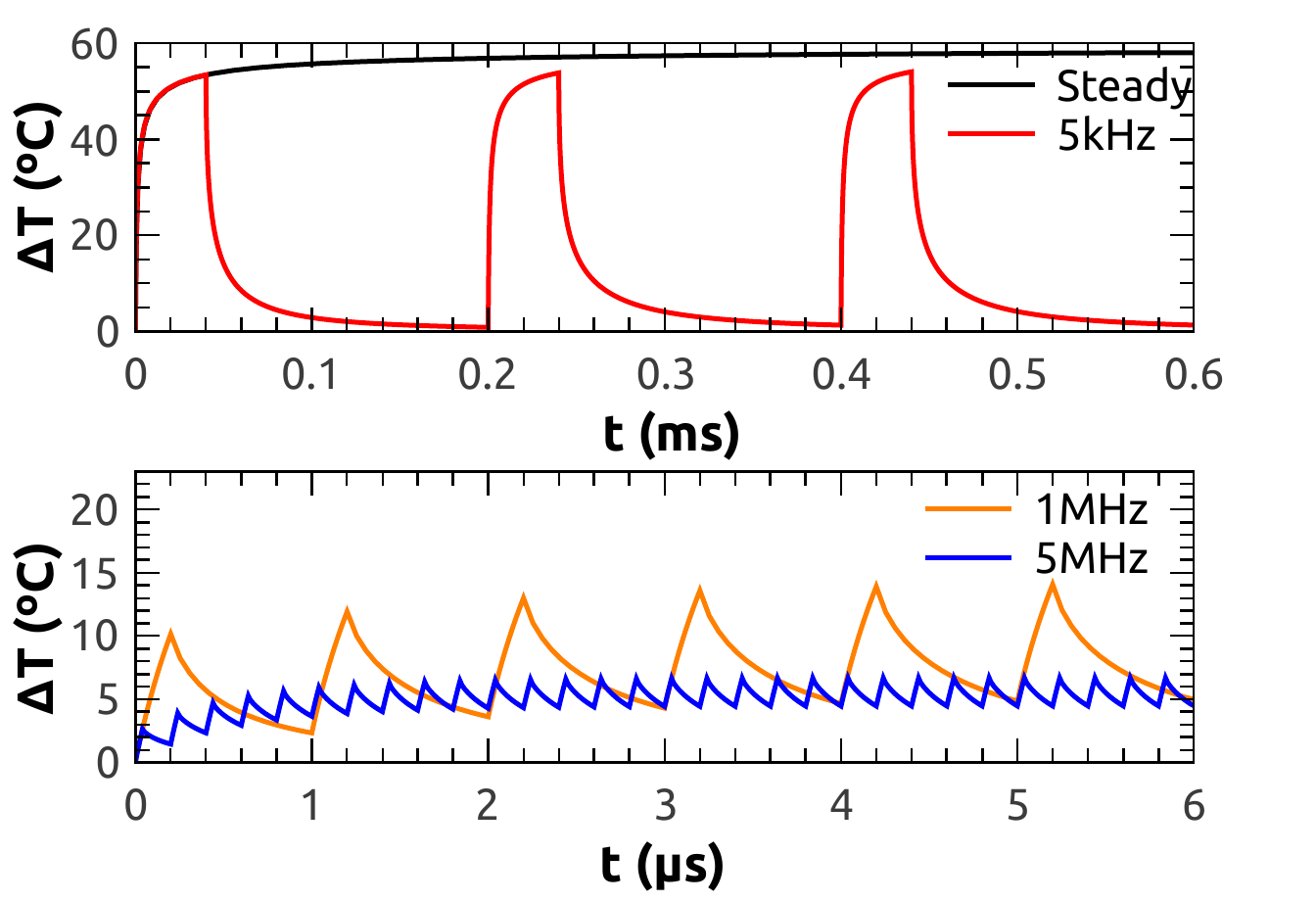}
\caption{Temporal evolution of the mean temperature change for the steady state condition (no modulation) and 3 different modulation frequencies, as calculated from \eqref{e:dif10} and \eqref{e:dif11}, for a 20\% duty cycle. \label{f:Tvst}}  
\end{figure*}}

The considered problem has two distinct regimes: $\xtx{q(t)}=\xtx{q}$ during the ON-cycle, and $\xtx{q(t)}=0$ during the OFF-cycle. Lets consider the ON-cycle first. The structure initially has an homogeneous temperature $\xtx{T}(\xtx{z},0)=\xtx{T}_0$. The solution to \ecref{e:dif2} has the following form \cite{DifT1}:
\begin{widetext}
\begin{align}
\xtx{T(x,y,z,t)} &= \xtx{T}_0 + \int_{0}^{t} \! \int_{-\infty}^{\infty} \! \int_{-\infty}^{\infty} \! \int_{-\infty}^{\infty} \! \frac{\xtx{Q(x',y',z',t')}}{\left( {4 \pi \alpha_T (\xtx{t}-\xtx{t'})} \right)^{3/2}} \xtx{e}^{\left( -\frac{(\xtx{x}-\xtx{x'})^2+(\xtx{y}-\xtx{y'})^2+(\xtx{z}-\xtx{z'})^2}{4 \alpha_T (\xtx{t}-\xtx{t'})} \right)} \xtx{dx'dy'dz'dt'} \nonumber \\
\xtx{T(x,y,z,t)} &= \xtx{T}_0 + \frac{\xtx{q}}{2} \sigma^2 \int_{0}^{t} \! \frac{e^{-\frac{\xtx{x}^2+\xtx{y}^2}{4 \alpha_T (\xtx{t}-\xtx{t'})+2 \sigma^2}}}{2 \alpha_T (\xtx{t}-\xtx{t'})+ \sigma^2} \cdot \sum e^{-\xtx{z}_i/\xtx{z}_0} {\left[ \Gamma \left( \frac{\xtx{z} - \xtx{z}_i - \xtx{d}_i}{\sqrt{4 \alpha_T (\xtx{t}-\xtx{t'})}} \right) - \Gamma \left( \frac{\xtx{z} - \xtx{z}_i}{\sqrt{4 \alpha_T (\xtx{t}-\xtx{t'})}} \right) \right]} \xtx{dt'} ,
\label{e:dif4}
\end{align}
\end{widetext}
where $\Gamma$ corresponds to the error function \cite{ErrorFunc}.

After the first ON-cycle, the temperature distribution will be given by
\begin{widetext}
\begin{equation}
\xtx{T(x,y,z,t)} = \xtx{T}_0 + \frac{\xtx{q}}{2} \sigma^2 \int_{0}^{\xtx{t}_{end}} \! \frac{e^{-\frac{\xtx{x}^2+\xtx{y}^2}{4 \alpha_T (\xtx{t}_{end}-\xtx{t'})+2 \sigma^2}}}{2 \alpha_T (\xtx{t}_{end}-\xtx{t'})+ \sigma^2} \cdot \sum e^{-\xtx{z}_i/\xtx{z}_0} {\left[ \Gamma \left( \frac{\xtx{z} - \xtx{z}_i - \xtx{d}_i}{\sqrt{4 \alpha_T (\xtx{t}_{end}-\xtx{t'})}} \right) - \Gamma \left( \frac{\xtx{z} - \xtx{z}_i}{\sqrt{4 \alpha_T (\xtx{t}_{end}-\xtx{t'})}} \right) \right]} \xtx{dt'},
\label{e:dif5}
\end{equation}
\end{widetext}
being $\xtx{t}_{end}$ the end time of the cycle. In the case when there is no heat source ($\xtx{q(t)}=0$), but there is a non homogeneous temperature profile, the solution to \ecref{e:dif2} is \cite{DifT1}
\begin{widetext}
\begin{equation}
\xtx{T(x,y,z,t)} = \int_{-\infty}^{\infty} \! \int_{-\infty}^{\infty} \! \int_{-\infty}^{\infty} \! \frac{\xtx{T} \xtx{(x',y',z',}\xtx{t}_{end})}{\left( {4 \pi \alpha_T (\xtx{t}-\xtx{t}_{end})} \right)^{3/2}} \xtx{e}^{\left( - \frac{ (\xtx{x}-\xtx{x'})^2+(\xtx{y}-\xtx{y'})^2+(\xtx{z}-\xtx{z'})^2}{4 \alpha_T (\xtx{t}-\xtx{t}_{end})} \right)} \xtx{dx'dy'dz'}.
\label{e:dif6}
\end{equation}
\end{widetext}
Comparing \ecref{e:dif4} and \ecref{e:dif6} it is evident that the latter is equal to the solution found for the problem of a heat source of the following form:
\begin{equation}
 \xtx{Q(x,y,z,t)} = \xtx{T} \xtx{(x,y,z,}\xtx{t}) \cdot \delta (\xtx{t}-\xtx{t}_{end}).
\label{e:dif7}
\end{equation}

Except for the very first ON-cycle, there is always a residual temperature distribution left from earlier cycles. To take this into account, we define the following effective heat source for each ON-cycle: 
\begin{widetext}
\begin{equation}
 \xtx{Q} \xtx{(x,y,z,t)} = \xtx{T} \xtx{(x,y,z,}\xtx{t}) \cdot \delta (\xtx{t}-\xtx{t}_{0,i}) + \xtx{q} \cdot \Theta (\xtx{t}-\xtx{t}_{0,i}) \cdot e^{-\frac{\xtx{x}^2+\xtx{y}^2}{2 \sigma^2}} \sum e^{-\xtx{z}_i/\xtx{z}_0} \bigg( \Theta \left( \xtx{z}_i \right) \Big( 1-\Theta \left( \xtx{z}_i+\xtx{d}_i \right) \Big) \bigg)  ,
\label{e:dif8}
\end{equation}
\end{widetext}
being $\xtx{t}_{0,i}$ the initial time of ON-cycle $i$ and $\Theta (\xtx{t}_{0,i})$ the Heaviside step function. During the OFF-cycles, on the other hand, we take
\begin{equation}
\xtx{Q} \xtx{(x,y,z,t)} = \xtx{T} \xtx{(x,y,z,}\xtx{t}) \cdot \delta (\xtx{t}-\xtx{t}_{end,i}),
\label{e:dif9}
\end{equation}
being $\xtx{t}_{end,i}$ the end time of the ON-cycle $i$. The solution of the problem will then be
\begin{widetext}
\begin{align}
\xtx{T(x,y,z,t)} =& \int_{-\infty}^{\infty} \! \int_{-\infty}^{\infty} \! \int_{-\infty}^{\infty} \! \frac{\xtx{T} \xtx{(x',y',z',}\xtx{t}_{0,i})}{\left( {4 \pi \alpha_T (\xtx{t}-\xtx{t}_{0,i})} \right)^{3/2}} \xtx{e}^{\left( - \frac{ (\xtx{x}-\xtx{x'})^2+(\xtx{y}-\xtx{y'})^2+(\xtx{z}-xtx{z'})^2}{4 \alpha_T (\xtx{t}-\xtx{t}_{0,i})} \right)} \xtx{dx'dy'dz'} \nonumber \\
&+ \frac{\xtx{q}}{2} \sigma^2 \int_{\xtx{t}_{0,i}}^{\xtx{t}} \! \frac{e^{-\frac{\xtx{x}^2+\xtx{y}^2}{4 \alpha_T (\xtx{t}-\xtx{t'})+2 \sigma^2}}}{2 \alpha_T (\xtx{t}-\xtx{t'})+ \sigma^2} \cdot \sum e^{-\xtx{z}_i/\xtx{z}_0} {\left[ \Gamma \left( \frac{\xtx{z} - \xtx{z}_i - \xtx{d}_i}{\sqrt{4 \alpha_T (\xtx{t}-\xtx{t'})}} \right) - \Gamma \left( \frac{\xtx{z} - \xtx{z}_i}{\sqrt{4 \alpha_T (\xtx{t}-\xtx{t'})}} \right) \right]} \xtx{dt'},
\label{e:dif10}
\end{align}
\end{widetext}
during the ON-cycles, and	
\begin{widetext}		
\begin{equation}
\xtx{T(x,y,z,t)} = \int_{-\infty}^{\infty} \! \int_{-\infty}^{\infty} \! \int_{-\infty}^{\infty} \! \frac{\xtx{T} \xtx{(x',y',z',}\xtx{t}_{end,i})}{\left( {4 \pi \alpha_T (\xtx{t}-\xtx{t}_{end,i})} \right)^{3/2}} \xtx{e}^{\left( - \frac{ (\xtx{x}-\xtx{x'})^2+(\xtx{y}-\xtx{y'})^2+(\xtx{z}-\xtx{z'})^2}{4 \alpha_T (\xtx{t}-\xtx{t}_{end,i})} \right)} \xtx{dx'dy'dz'}.
\label{e:dif11}
\end{equation}
\end{widetext}
during the OFF-cycles.

\begin{figure*}[hhht]
\centering{}\includegraphics[width=120mm]{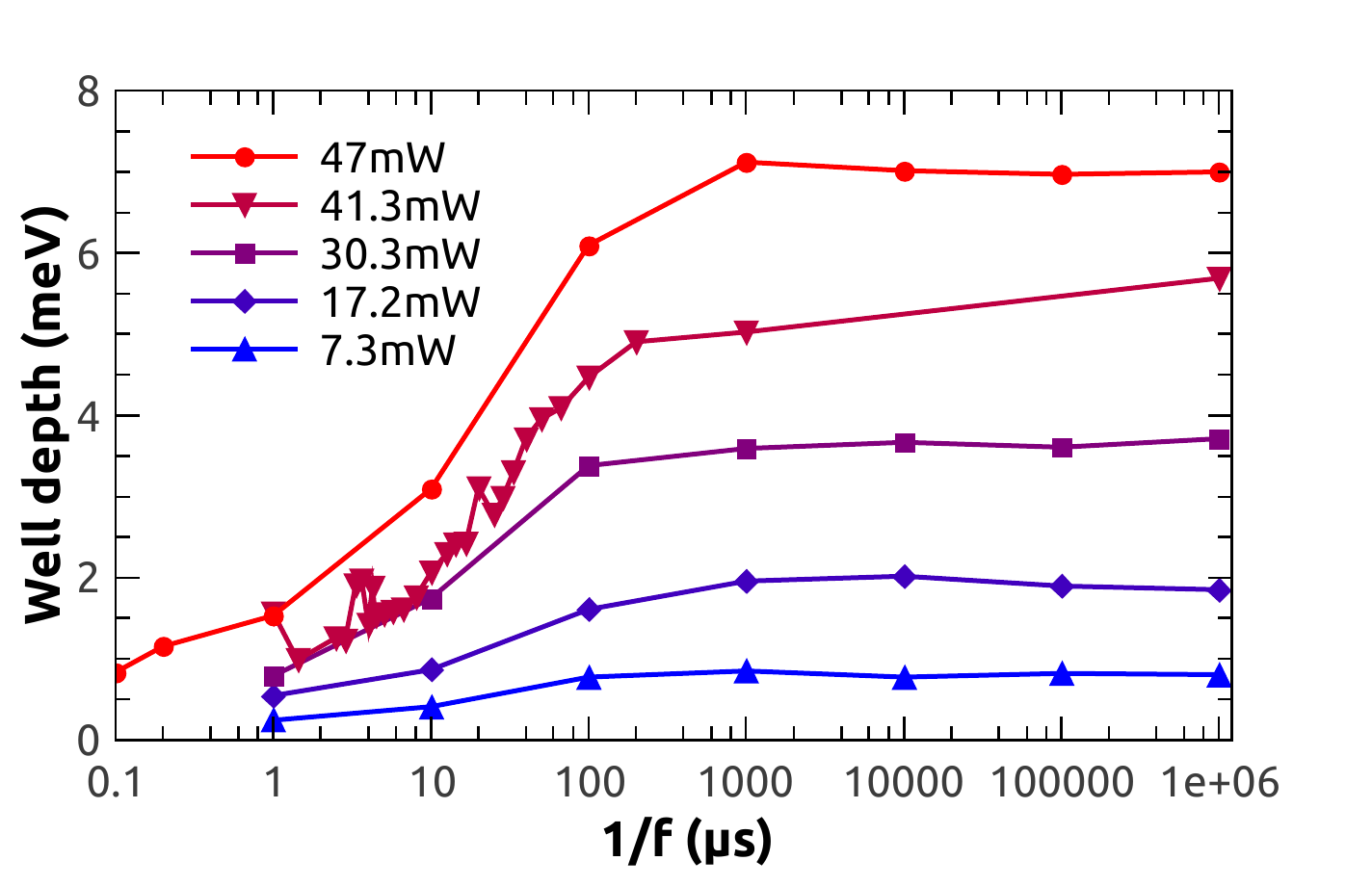}
\caption{Optical well depth as a function of the modulation period for excitation powers ranging from 7.3\mW \ to 47\mW \ and a duty cycle of 20{\%}. \label{f:vsfrecyP}}  
\end{figure*}

After a certain number of cycles, which will depend on the ratio between the excitation time ($\xtx{t}_{on}$) and the recovery time ($\xtx{t}_{off}$), a quasi-stationary state is accomplished. From this point forwards, the sample will reach the same temperature distribution at each cycle. The maximum temperature that will be reached will therefore depend on both the period and the duty cycle of the modulation. An example of the resulting temporal evolution of the structure's mean temperature change for 3 different modulation frequencies, as well as the steady state, are presented in \figref{f:Tvst}. The thermal diffusivities used in the simulations were taken from the literature and are summarized in \tableref{t:1}.

The temperature of the structure is, as already mentioned, proportional to the optical mode shift (which in turn is proportional to the change in index of refraction of the materials), and thus proportional to the optical potential well depth. From photoluminescence measurements performed as a function of temperature, we found that above 290\K \ and up to at least 340\K , the spectral position of the optical mode depends linearly with temperature with a slope of $\frac{\partial \varepsilon}{\partial T} = 0.118 \pm 0.004$\,$\frac{\xtx{meV}}{\xtx{K}}$. Therefore, a \simi 2.1\meV \ optical well as the one presented in \figref{f:teoria} corresponds to a radial difference in temperature of \simi 18\K .

\begin{table}[]
\caption{Thermal diffusivity}
\centering
\begin{tabular}{ l l}
\specialrule{.15em}{.075em}{.075em}
\textbf{Medium} & $\alpha_T$ $\left( \frac{\mu \xtx{m}^2}{\mu \xtx{s}} \right)$  \\ 
\specialrule{.15em}{.075em}{.075em}
\textbf{GaAs} \cite{Carlson,Blakemore}									     & $31.0$   \\
$\mathbf{\xtx{\textbf{Al}}_{0.1}\xtx{\textbf{Ga}}_{0.9}\xtx{\textbf{As}}}$ \cite{Afromowitz,Lichter}   	 & $20.0$   \\
$\mathbf{\xtx{\textbf{Al}}_{0.95}\xtx{\textbf{Ga}}_{0.05}\xtx{\textbf{As}}}$ \cite{Afromowitz,Lichter}    & $54.0$    \\
\textbf{Air} \cite{Shen,Mandelis}									     & $20.0$    \\
\specialrule{.075em}{0.075em}{0em}
\end {tabular}
\label{t:1}
\end{table}

\section{Frequency dependence of the optical well}
In Fig. 5 of the main text, we presented several photoluminescence measurements, performed for different modulation frequencies, and a fixed duty cycle of 20\%. From red to blue, this frequencies are: 1\Hz , 1\kHz , 5\kHz , 10\kHz , 15\kHz , 20\kHz , 25\kHz , 30\kHz , 35\kHz , 40\kHz , 50\kHz , 60\kHz , 70\kHz , 80\kHz , 100\kHz , 125\kHz , 150\kHz , 175\kHz , 200\kHz , 225\kHz , 235\kHz , 250\kHz , 275\kHz , 300\kHz , 350\kHz , 400\kHz , 700\kHz \ and 1\MHz . The corresponding optical well depths are plotted in \figref{f:vsfrecyP} (41.3\mW), alongside the ones already shown in Fig. 5 of the main text (47\mW) and 3 more sets of measurements corresponding to different excitation powers. It is evident from this figure that the saturation above \simi 1kHz (1/f = 1000\mus) is a consistent feature, present for every excitation power used.

\end{document}